\documentclass[12pt]{article}
\usepackage{amssymb, amsthm}
\usepackage{amsmath}    
\usepackage{mathrsfs}   
\usepackage{graphicx}   
\usepackage{color}      
\usepackage{footnote}   
\makesavenoteenv{tabular}
\makesavenoteenv{table}
\usepackage{slashed}    
\usepackage{subfigure}  
\usepackage{hyperref}   
\usepackage{authblk}    
\usepackage{rotating}   
\usepackage{multirow}   

\usepackage{fancyhdr}
\fancypagestyle{plain}{
        \fancyhead[R]{\fbox{NT@UW-15-06}}
        }

\begin{document}

\title{Polarized Lepton-Nucleon Elastic Scattering and a Search for a Light Scalar Boson}
\author[]{Yu-Sheng Liu\footnote{ysliu@uw.edu}, and Gerald A. Miller\footnote{miller@uw.edu}}
\affil[]{Department of Physics, University of Washington, Seattle, WA 98195-1560, USA}

\maketitle

\begin{abstract}
Lepton-nucleon elastic scattering, using the one-photon and one-scalar-boson exchange mechanisms considering all possible polarizations, is used to study searches for a new scalar boson and suggest new measurements of the nucleon form factors. A new light scalar boson, which feebly couples to leptons and nucleons, may account for the proton radius and muon $g-2$ puzzles. We show that the scalar boson produces relatively large effects in certain kinematic region when using sufficient control of lepton and nucleon spin polarization. We generalize current techniques to measure the ratio $G_E/G_M$ and present a new method to separately measure $G_M^2$ and $G_E^2$ using polarized incoming and outgoing muons.
\end{abstract}

\section{Introduction}
Lepton-nucleon elastic scattering is important both in theory and experiment. The use of polarization techniques has yielded much new information. In this paper, we study lepton-nucleon elastic scattering using the one-photon and one-scalar-boson exchange mechanism by considering all possible polarizations. The differential cross sections are calculated in a general reference frame. There are two applications of our results in this paper. The first one is searching for a new light scalar boson. The second one, using only the one-photon exchange contribution, is to provide new ways to 
measure the nucleon form factor: generalizing current techniques to measure the ratio $G_E/G_M$ and presenting a new method to separately measure $G_M^2$ and $G_E^2$ using polarized incoming and outgoing muons.

The interest in a new light scalar boson arises from recent studies of the proton radius using the Lamb shift in the $2{\rm S}_{1/2}-2{\rm P}_{3/2}$ transition in muonic hydrogen \cite{Pohl:2010zza,Antognini:1900ns}, the value of proton radius was reported to be 0.84087(39) fm, whereas the CODATA value \cite{Mohr:2012tt} 0.8775(51) fm is 7 standard deviation away. The major difference between these two reported data is that the former extracts the proton radius from muonic hydrogen and the latter from electronic hydrogen and electron-proton scattering experiments. Although the different proton radius may arise due to subtle lepton-nucleon non-perturbative effects within the standard model \cite{Pohl:2013yb}, it could also be a signal of new physics caused by a violation of lepton universality.

Another candidate of new physics signal is the muon anomalous magnetic moment, which is defined as
$a_\mu=\frac{(g-2)_\mu}{2}$. The measurement at BNL \cite{Blum:2013xva} differs from the standard model prediction by 3 to 4 standard deviations,\footnote{The experimental and the standard model uncertainty are $63\times 10^{-11}$ and $\simeq 49\times 10^{-11}$, respectively.}
\begin{align}
\Delta a_\mu=a_\mu^{\rm exp}-a_\mu^{\rm th}=&(287\pm 80)\times 10^{-11}\;\cite{Davier:2010nc}\\
=&(261\pm 78)\times 10^{-11}\;\cite{Hagiwara:2011af}
\end{align}
the different values depending on the choice of lowest order hadronic contribution. This discrepancy is also possibly explained by new physics involving violation of lepton universality.

It is known that a light scalar boson with mass around 1 MeV is a candidate to explain both proton radius and muon anomalous magnetic moment puzzles simultaneously \cite{Pohl:2013yb,TuckerSmith:2010ra,Izaguirre:2014cza}. The non-relativistic potential between lepton and nucleon caused by exchange of a scalar boson is written as
\begin{align}
V_\phi(r)=-\frac{g_lg_N}{4\pi}\frac{e^{-m_\phi r}}{r}=\frac{g_lg_N}{e_le_N}V_{EM}(r)e^{-m_\phi r},
\end{align}
where $m_\phi$ is the mass of the scalar boson; $g_l$ and $g_N$ are scalar bosons coupling to lepton and nucleon, respectively. We adopt the constraints in \cite{TuckerSmith:2010ra} that for $m_\phi=1$ MeV
\begin{align}\label{eq:constraints on g}
\frac{g_e}{e}\lesssim 2.3\times10^{-4},\;\frac{g_p}{e}=\frac{g_\mu}{e}\lesssim 1.3\times10^{-3},\;g_n\lesssim 2\times10^{-5}.
\end{align}
So the potential of the scalar boson is suppressed by $\frac{g_l}{e_l}\frac{g_N}{e_N}$ and an exponentially decay factor comparing with the Coulomb potential. The fact that the potential of the scalar boson is intrinsically much smaller than the Coulomb potential makes it hard to find the scalar boson. There could be other possible particles, however, most of them are ruled out, e.g. pseudoscalar and axial vector are ruled out by hyperfine splittings \cite{Pohl:2013yb,Carlson:2012pc}.

There are proposals to search for the light scalar boson, such as a direct detection method \cite{Izaguirre:2014cza}. In this paper, we study the cross section of elastic lepton-nucleon scattering caused by one-photon and one-scalar-boson exchange. Since the muon is much heavier than the electron, the lepton mass can not be neglected, as is done for electron-proton scattering. The two-photon exchange contribution is expected (from perturbation expansion) and measured to be at a few percent level compared with one-photon exchange (OPE) \cite{Blunden:2005ew,Afanasev:2005mp,Arrington:2007ux,Carlson:2007sp}. Therefore, the effect of one-scalar-boson exchange must be greater than few percent of OPE to be observed. We consider unpolarized and polarized elastic scattering cross sections. We find the following two cases that for certain kinematic regions where the effects of the scalar boson are dominant and potentially observable: electron-neutron cross section with incoming and outgoing electrons polarized, and muon-neutron cross section with incoming and outgoing neutrons polarized. 

There are several facilities that can measure the lepton-nucleon elastic scattering, for electron, such as JLab, and for muon, MUSE at PSI and the J-PARC muon facility. Based on our results and the current experimental setup and capability \cite{SoLID:2013fta,MUSE}, although the polar angle resolution is sufficient, the kinematic region where scalar bosons may have significant effects is beyond the current polar angle measured range; the electron-neutron scattering is more promising than muon-neutron scattering due to the high intensity of the electron flux. Further estimates and discussions are in section \ref{sec:searching for phi_leptons polarized} and \ref{sec:searching for phi_nucleons polarized}.

The polarized cross section with one-photon exchange is used to measure the Sachs form factors, $G_M$ and $G_E$ \cite{Rock:2001zi,Arrington:2011kb,Punjabi:2014tna}. The standard technique of measurement is using elastic electron-nucleon scattering using the following experiments: unpolarized Rosenbluth separation \cite{Rosenbluth:1950yq}, polarized lepton beam and nucleon target, and polarized lepton beam and recoil nucleon. The latter two experiments, using the ratio technique (also known as polarization transfer method), were first developed in \cite{Akhiezer:1974em} and later discussed in more detail in \cite{Arnold:1980zj}. We explore other possible ways to determine form factors by including lepton masses and other lepton polarization configurations different from conventional ones. In section \ref{sec:form factor ij}, we generalize the current method to measure the ratio of form factors, $G_E/G_M$, by including lepton mass, non-longitudinal lepton polarization, and more polarization configurations (polarized one lepton and one nucleon, either they are incoming or outgoing). In section \ref{sec:form factor 13}, we present a new method to measure $G_E^2$ or $G_M^2$ directly for certain kinematic conditions in elastic muon-nucleon scattering cross section with polarized incoming and outgoing muons.

The outline of this paper is as follows. In section \ref{sec:setup}, we set up the formalism needed to compute cross sections. In section \ref{sec:cross section}, the cross sections are calculated in a general reference frame with all possible polarizations, and the massless lepton limit is slso discussed. In section \ref{sec:searching for a scalar boson}, we show that there are two cases in which the scalar boson is potentially observable: electron-neutron cross section with incoming and outgoing electrons polarized, and muon-neutron cross section with incoming and outgoing neutrons polarized. In section \ref{sec:measuring form factors}, comparing with the current method, we give a more general result for the ratio of form factors, $G_E/G_M$, by including lepton mass, non-longitudinal lepton polarization, and more polarization configurations. We also discuss new measurements to selectively obtain the contribution from $G_E^2$ or $G_M^2$ in the cross section with polarized incoming and outgoing muons. A conclusion is presented in section \ref{sec:conclusions}.

\section{Setup\label{sec:setup}}
The elastic lepton-nucleon scattering process is denoted as
\begin{align}
l(p_1,s_1)+N(p_2,s_2)\rightarrow l(p_3,s_3)+N(p_4,s_4),
\end{align}
where $l$ and $N$ stand for lepton and nucleon; $p$ and $s$ are momentum and spin polarization; the number 1, 2, 3, and 4 label the incoming lepton, incoming nucleon, outgoing lepton, and outgoing nucleon. This notation is used throughout the entire paper. In the lowest order, we consider the interaction by exchanging a scalar boson or a photon between the lepton and the nucleon. There are three contributions to the cross section: one-photon exchange, one-scalar-boson exchange, and the interference terms.

\subsection{Kinematics}
In the nucleon rest frame (lab frame), we choose the the coordinate such that $\mathbf{p}_1$ is along $z$ axis and $\mathbf{p}_3$ is in $x$-$z$ plane to exploit the symmetry. With these choices, all the kinematics quantities which we need can be expressed in terms of only two variables, $|\mathbf{p}_1|$ and the scattering angle $\theta$. The scattering angle is an angle between outgoing and incoming lepton, and defined as $\cos\theta=\hat{\mathbf{p}}_1\cdotp\hat{\mathbf{p}}_3$. In the lab frame, the external momenta and space-like momentum transfer $q$ can be expressed as follows
\begin{align}
\mathbf{p}_1&=|\mathbf{p}_1|\hat{\mathbf{z}},\; E_1=\sqrt{|\mathbf{p}_1|^2+m_l^2},\;\mathbf{p}_2=\mathbf{0},\; E_2=m_N,\\
\mathbf{p}_3&=\frac{(m_l^2+E_1m_N)\cos\theta+(E_1+m_N)\sqrt{m_N^2-m_l^2\sin^2\theta}}{(E_1+m_N)^2-\cos^2\theta|\mathbf{p}_1|^2}|\mathbf{p}_1|\hat{\mathbf{p}_3},\\
\hat{\mathbf{p}_3}&=\sin\theta\hat{\mathbf{x}}+\cos\theta\hat{\mathbf{z}},\\
E_3&=\frac{(E_1+m_N)(m_l^2+E_1m_N)+\cos\theta|\mathbf{p}_1|^2\sqrt{m_N^2-m_l^2\sin^2\theta}}{(E_1+m_N)^2-\cos^2\theta|\mathbf{p}_1|^2}\\
\mathbf{p}_4&=\mathbf{p}_1-\mathbf{p}_3,\; E_4=E_1+m_N-E_3,\\
q&=p_1-p_3=p_4-p_2,\\
q^2&=\frac{2|\mathbf{p}_1|^2m_N(m_N+E_1\sin^2\theta-\cos\theta\sqrt{m_N^2-m_l^2\sin^2\theta})}{(E_1+m_N)^2-\cos^2\theta|\mathbf{p}_1|^2}.
\end{align}
We use the mostly-plus metric, so that $q^2>0$ if $q$ is space-like. If we take spin polarization into account, each polarized particle needs two angular variables to specify the spin direction, see section \ref{sec:polarization} below.

\subsection{Dynamics}
\begin{figure}
\centering
\includegraphics[scale=0.8]{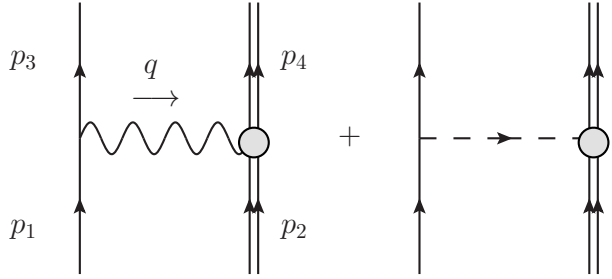}
\caption{\label{fig:amplitude} The tree level amplitudes of lepton-nucleon elastic scattering: the single (double) line on the left (right) denotes lepton (nucleon); the wavy (dashed) line denotes photon (scalar boson).}
\end{figure}

The scalar boson interacts with lepton and nucleon through Yukawa coupling
\begin{align}
\mathcal{L}_\phi\supset-\frac{1}{2}(\partial\phi)^2-\frac{1}{2}m_\phi^2\phi^2+g_l\phi\bar{\psi}_l\psi_l+g_N\phi\bar{\psi}_N\psi_N,
\end{align}
where $g_l$ and $g_N$ are Yukawa couplings between lepton-scalar and nucleon-scalar. There are two lowest order diagrams: one-photon and one-scalar-boson exchange, see figure \ref{fig:amplitude}.

The amplitude squared is given by
\begin{align}\label{eq:amplitudes}
|\mathcal{T}|^2=e_l^2 e_N^2\left(I_{\gamma\gamma}+\lambda I_{\gamma\phi}+\lambda^2I_{\phi\phi}\right),
\end{align}
where $e_l$ and $e_N$ are $U(1)_{EM}$ couplings of leptons and nucleons, respectively; $\lambda$ and $I_{xy}$ using the mostly-plus metric are given by
\begin{align}
\lambda(q^2)=&\frac{g_lg_N}{e_le_N}\frac{q^2}{q^2+m_\phi^2}\label{eq:lambda}\\
I_{\gamma\gamma}=&\frac{1}{q^4}{\rm Tr}(u_3\bar{u}_3\gamma_\mu u_1\bar{u}_1\gamma_\nu){\rm Tr}(u_4\bar{u}_4 V^\mu u_2\bar{u}_2 V^\nu)\\
I_{\gamma\phi}=&\frac{1}{q^4}\Big[{\rm Tr}(u_3\bar{u}_3\gamma_\mu u_1\bar{u}_1){\rm Tr}(u_4\bar{u}_4 V^\mu u_2\bar{u}_2)\nonumber\\
&\quad+{\rm Tr}(u_3\bar{u}_3 u_1\bar{u}_1\gamma_\mu){\rm Tr}(u_4\bar{u}_4 u_2\bar{u}_2 V^\mu)\Big]\\
I_{\phi\phi}=&\frac{1}{q^4}{\rm Tr}(u_3\bar{u}_3 u_1\bar{u}_1){\rm Tr}(u_4\bar{u}_4 u_2\bar{u}_2)\\
V^\mu(q)=&F_1(q^2)\gamma^\mu-\frac{i}{2m_N}F_2(q^2)\sigma^{\mu\nu}q_\nu
\end{align}
where $m_\phi$ is mass of the scalar boson; $\sigma^{\mu\nu}=\frac{i}{2}[\gamma^\mu,\gamma^\nu]$; $F_1(q^2)$ and $F_2(q^2)$ are the form factors of nucleon $U(1)_{EM}$ coupling. In general, there is a form factor, $F_3(q^2)$, of scalar-nucleon coupling, however, it always appears with $g_N$, therefore we can include it into the constraint of $g_N$.

\subsection{Polarization \label{sec:polarization}}
The product of spinors $u\bar{u}$ in mostly-plus metric is given by
\begin{align}\label{eq:uu}
u\bar{u}=\frac{1}{2}(1-\gamma_5\slashed{s})(-\slashed{p}+m)
\end{align}
where $s$ is the spin polarization which satisfies $s\cdotp p=0$ and $s^2=1$. We can solve for the spin polarization using the two constraints
\begin{align}
s=z\frac{E}{\sqrt{m^2+|\mathbf{p}|^2\sin^2\alpha}}\left(\frac{|\mathbf{p}|}{E}\cos\alpha,\hat{\mathbf{s}}\right),
\end{align}
where $\alpha$ and $\beta$ are the polar and azimuthal angle of $\mathbf{s}$ with respect to $\mathbf{p}$; $\cos\alpha=\hat{\mathbf{p}}\cdotp\hat{\mathbf{s}}$; $z=\pm 1$ and becomes helicity if the particle is longitudinally polarized.

The polarization of leptons and nucleons in the lab frame in the coordinate we chose can be expressed as
\begin{align}
\hat{\mathbf{s}}_i=&\sin\alpha_i\cos\beta_i\hat{\mathbf{x}}+\sin\alpha_i\sin\beta_i\hat{\mathbf{y}}+\cos\alpha_i\hat{\mathbf{z}}\quad\text{for $i=$1 or 2}\\
\hat{\mathbf{s}}_3=&\sin\alpha_3\cos\beta_3(\hat{\mathbf{y}}\times\hat{\mathbf{p}}_3)+\sin\alpha_3\sin\beta_3\hat{\mathbf{y}}+\cos\alpha_3\hat{\mathbf{p}}_3\nonumber\\
=&(\sin\alpha_3\cos\beta_3\cos\theta+\cos\alpha_3\sin\theta)\hat{\mathbf{x}}+\sin\alpha_3\sin\beta_3\hat{\mathbf{y}}\nonumber\\
&+(-\sin\alpha_3\cos\beta_3\sin\theta+\cos\alpha_3\cos\theta)\hat{\mathbf{\mathbf{z}}}\\
\hat{\mathbf{s}}_4=&(\sin\alpha_4\cos\beta_4\cos\theta'-\cos\alpha_4\sin\theta')\hat{\mathbf{x}}+\sin\alpha_4\sin\beta_4\hat{\mathbf{y}}\nonumber\\
&+(\sin\alpha_4\cos\beta_4\sin\theta'+\cos\alpha_4\cos\theta')\hat{\mathbf{\mathbf{z}}}
\end{align}
where $\theta'$ is angle between $\mathbf{p}_4$ and $\mathbf{p}_1$; $\cos\theta'=\hat{\mathbf{p}}_4\cdotp\hat{\mathbf{p}}_1=\frac{|\mathbf{p}_1|-|\mathbf{p}_3|\cos\theta}{\sqrt{E_4^2-m_N^2}}$.

There are two important special cases. For transverse polarization $\alpha_T=\frac{\pi}{2}$, $s_T=z(0,\hat{\mathbf{s}})$. For longitudinal polarization $\alpha_L=0$,
\begin{align}\label{eq:sL}
s_L=z\frac{E}{m}\left(\frac{|\mathbf{p}|}{E},\hat{\mathbf{p}}\right).
\end{align}
This is useful for a light lepton because if its mass is small compared with the beam energy, the light lepton is naturally longitudinally polarized. 

\subsubsection{Massless Particle \label{sec:polarization of massless particle}}
The massless lepton limit is important for electron scattering or the high energy limit. In general, naively taking $m_l\rightarrow 0$ in physical quantities does not give us the correct result (see section \ref{sec:massless lepton} below), because the longitudinal polarization (\ref{eq:sL}) blows up. The correct way to take the limit is to set the polarization to be longitudinal, $s_\mu\rightarrow p_\mu/m_l$, then take the massless limit. Then the product of spinors (\ref{eq:uu}) becomes
\begin{align}
u\bar{u}=-\frac{1}{2}(1+z\gamma_5)\slashed{p}{\rm\; (massless)},
\end{align}
where $z=\pm 1$ is the helicity of the lepton.

\section{Cross Section\label{sec:cross section}}
Combining with the amplitude squared (\ref{eq:amplitudes}), the differential cross section in a general reference frame is
\begin{align}\label{eq:cross section}
\frac{d\sigma}{d\Omega}=A(I_{\gamma\gamma}+\lambda I_{\gamma\phi}+\lambda^2 I_{\phi\phi})=\left(\frac{d\sigma}{d\Omega}\right)_{\gamma\gamma}+\left(\frac{d\sigma}{d\Omega}\right)_{\gamma\phi}+\left(\frac{d\sigma}{d\Omega}\right)_{\phi\phi},
\end{align}
where
\begin{align}
A=\frac{\alpha^2}{4}\frac{|\mathbf{p}_3|}{|E_2\mathbf{p}_1-E_1\mathbf{p}_2|}\left[E_1+E_2-E_3\frac{(\mathbf{p}_1+\mathbf{p}_2)\cdotp\hat{\mathbf{p}}_3}{|\mathbf{p}_3|}\right]^{-1}.
\end{align}
In the lab frame, the kinematic factor $A$ becomes
\begin{align}
A_{\rm lab}=\frac{\alpha^2}{4m_N}\frac{|\mathbf{p}_3|}{|\mathbf{p}_1|}\left[E_1+m_N-E_3\frac{|\mathbf{p}_1|}{|\mathbf{p}_3|}\cos\theta\right]^{-1}.
\end{align}

In (\ref{eq:cross section}), it is worth mentioning that the pure one-photon or one-scalar-boson exchange contribution, $\left(\frac{d\sigma}{d\Omega}\right)_{\gamma\gamma}$ and $\left(\frac{d\sigma}{d\Omega}\right)_{\phi\phi}$, is always greater or equal to zero, whereas the interference term $\left(\frac{d\sigma}{d\Omega}\right)_{\gamma\phi}$ can be positive or negative depending on the sign of Yukawa coupling $g$ and electric charge.

\subsection{Unpolarized}
The cross section (\ref{eq:cross section}) is proportional to $I_{\gamma\gamma}$, $I_{\gamma\phi}$, and $I_{\phi\phi}$, so it is sufficient to show these three $I$'s instead of full cross section. The unpolarized $I$'s are defined as
\begin{align}
I^{\rm u}=\left(\frac{1}{2}\sum_{z_1}\right)\left(\frac{1}{2}\sum_{z_2}\right)\sum_{z_3}\sum_{z_4}I^{1,2,3,4}
\end{align}
where the superscripts refer to which particle is polarized and u for unpolarized. In a general reference frame,
\begin{align}
&I^{\rm u}_{\gamma\gamma}=\frac{G_E^2+\tau G_M^2}{1+\tau}\left(R^2-\frac{1+\tau}{\tau}\right)+G_M^2\left(2-\frac{m_l^2}{\tau m_N^2}\right)\\
&I^{\rm u}_{\gamma\phi}=-2G_E R\frac{m_l}{\tau m_N}\label{eq:I_gp unpol}\\
&I^{\rm u}_{\phi\phi}=\frac{1+\tau}{\tau}\left(1+\frac{m_l^2}{\tau m_N^2}\right)
\end{align}
where $G_M$ and $G_E$ are Sachs form factors
\begin{align}
G_M(q^2)=F_1(q^2)+F_2(q^2),\; G_E(q^2)=F_1(q^2)-\tau F_2(q^2);
\end{align}
$\tau=\frac{q^2}{4m_N^2}$ is defined in the usual way\footnote{Note that there is no minus sign in mostly-plus metric, and $\tau>0$ if $q$ is space-like.}; $R$ depends on Mandelstam variables\footnote{$s=-(p_1+p_2)^2$, $t=-(p_1-p_3)^2=-q^2$, and $u=-(p_1-p_4)^2$.}
\begin{align}
R=\frac{u-s}{t}=\frac{(p_1+p_3)\cdotp p_2}{(p_1-p_3)\cdotp p_2}=-\frac{4 p_1\cdotp p_2}{q^2}-1.
\end{align}
As an example, in the lab frame, $R_{\rm lab}=\frac{E_1+E_3}{E1-E_3}$, with massless lepton limit ($m_l\rightarrow0$), the Rosenbluth cross section is 
\begin{align}
\left(\frac{d\sigma}{d\Omega}\right)_{\rm Rosenbluth}=\left(AI^{\rm u}_{\gamma\gamma}\right)_{{\rm lab},m_l\rightarrow0}.
\end{align}

The interference term (\ref{eq:I_gp unpol}) is proportional to lepton mass $m_l$. The interference term for muon is about two order of magnitude bigger than for electron, see figure \ref{fig:unpolarized cross section}.

\subsection{Partially Polarized\label{sec:partially polarized}}
There is no contribution to cross section if only one particle is polarized in exchanging one-photon and one-scalar-boson. The reason is the following. The parity flips momentum and time reversal flips momentum and angular momentum. One can only flip spin polarization by combining parity and time reversal, and it is equivalent to change the overall sign of the polarization. Therefore, if the theory conserves $PT$, the spin asymmetry part can depend only on product of even number of spin polarizations to remain invariant under $PT$. For example, if we polarize three particles, there is no term depending on the product of all three polarizations, but there are terms depending on product of two polarizations, see section \ref{sec:fully polarized}. On the other hand, one can consider an interaction which breaks time reversal to have an additional dependence on odd number of spin polarizations, such as exchanging a Z boson. In conclusion, in exchanging a photon and a scalar boson, one needs to polarize at least two particles to have a spin dependent part.

\subsubsection{One Lepton and One Nucleon Polarized}
First we study the case that one lepton and one nucleon are polarized (each lepton and nucleon can be incoming or outgoing). In order to combine all four cases into one expression $I^{i,j}$, we require that the first superscript, $i$, to be lepton (1 or 3); the second superscript, $j$, to be nucleon (2 or 4).
\begin{align}
I^{i,j}=a^{i,j}\sum_{z_{k}}\sum_{z_{l}}I^{i,j,k,l}=a^{i,j}\Big[I^{\rm u}+s_\mu^i s_\nu^j(J^{i,j})^{\mu\nu}\Big]
\end{align}
where $I^{\rm u}$ is the unpolarized part; $i$, $j$ are particle labels and not summed; $a^{i,j}$ takes care of the factors $\frac{1}{2}$ due to the difference between average or sum over initial or final states
\begin{align}
a^{i,j}=\begin{cases}
1 & \text{if }(i,j)=(1,2)\\
\frac{1}{2} & \text{if }(i,j)=(1,4)\text{ or }(3,2)\\
\frac{1}{4} & \text{if }(i,j)=(3,4)\\
\end{cases} 
\end{align}
or more compactly $a^{i,j}=\left(\frac{1}{2}\right)^{\frac{i+j-3}{2}}$. The results are 
\begin{align}
(J_{\gamma\gamma}^{i,j})^{\mu\nu}=&2\frac{m_l}{m_N}G_M\Bigg[b^j F_2\frac{(p_2+p_4)^\mu q^\nu}{q^2}-\frac{G_E}{\tau}\Bigg(g^{\mu\nu}-\frac{q^\mu q^\nu}{q^2}\Bigg)\Bigg]\label{eq:I_gg^ij 1}\\
(J_{\gamma\phi}^{i,j})^{\mu\nu}=&2G_M\Bigg[R\Bigg(g^{\mu\nu}-\frac{q^\mu q^\nu}{q^2}\Bigg)+\frac{(p_2+p_4)^\mu(p_1+p_3)^\nu}{q^2}\Bigg]\\
(J_{\phi\phi}^{i,j})^{\mu\nu}=&0
\end{align}
where $b^j$ depends only on nucleon label
\begin{align}
b^j=\begin{cases}
1 & \text{if } j=2\\
-1 & \text{if } j=4\\
\end{cases} 
\end{align}
or more compactly $b^j=(-1)^{\frac{j}{2}+1}$. The expression of $(J_{\gamma\gamma}^{i,j})^{\mu\nu}$ (\ref{eq:I_gg^ij 1}) is the most compact form we found. However, the following expression for $(J_{\gamma\gamma}^{i,j})^{\mu\nu}$
\begin{align}\label{eq:I_gg^ij 2nd}
(J_{\gamma\gamma}^{i,j})^{\mu\nu}=2\frac{m_l}{m_N}\frac{G_M}{1+\tau}\Bigg[&b^j G_M\frac{(p_2+p_4)^\mu q^\nu}{q^2}-G_E\Bigg(2b^j \frac{p_j^\mu q^\nu}{q^2}+\frac{1+\tau}{\tau}g^{\mu\nu}-\frac{q^\mu q^\nu}{\tau q^2}\Bigg)\Bigg]
\end{align}
is more useful for measuring the Sachs form factors, $G_M$ and $G_E$, by polarized method, which is discussed in section \ref{sec:measuring form factors}.

It is worth noticing that it seems that in the massless lepton limit $(J_{\gamma\gamma}^{i,j})^{\mu\nu}$ vanishes, however it is not true because there is another $m_l$ in the denominator when the lepton spin polarization becomes longitudinal (\ref{eq:sL}). Therefore, this expression is well-behaved in the massless limit. We discuss the massless limit in detail in section \ref{sec:massless lepton}.

\subsubsection{Incoming and Outgoing Leptons Polarized}
For polarized incoming and outgoing leptons, the observable quantity is
\begin{align}
I^{1,3}=\left(\frac{1}{2}\sum_{z_2}\right)\sum_{z_4}I^{1,2,3,4}=\frac{1}{2}(1+s^1\cdotp s^3)I^{\rm u}+s_\mu^1 s_\nu^3(J^{1,3})^{\mu\nu}.
\end{align}
The results are
\begin{align}
(J_{\gamma\gamma}^{1,3})^{\mu\nu}=&-G_M^2 g^{\mu\nu}+2\frac{G_E^2+\tau G_M^2}{1+\tau}\Bigg(\frac{q^\mu q^\nu}{2\tau q^2}-\frac{p_2^{\{\mu} p_4^{\nu\}}+R q_{\phantom{1}}^{[\mu} p_2^{\nu]}}{q^2}\Bigg)\\
(J_{\gamma\phi}^{1,3})^{\mu\nu}=&2\frac{m_l}{\tau m_N}G_E \frac{q_{\phantom{1}}^{[\mu} p_2^{\nu]}}{q^2}\\
(J_{\phi\phi}^{1,3})^{\mu\nu}=&-\frac{1+\tau}{\tau} \frac{q^\mu q^\nu}{q^2}
\end{align}
where the curly and square brackets are symmetrization and anti-symmetrization notation
\begin{align}
p^{\{\mu}q^{\nu\}}=p^\mu q^\nu+p^\nu q^\mu,\; p^{[\mu}q^{\nu]}=p^\mu q^\nu-p^\nu q^\mu.
\end{align}
The expression in terms of $(J_{\gamma\gamma}^{1,3})^{\mu\nu}$ is the most compact form we found. The following expression of $I_{\gamma\gamma}^{1,3}$
\begin{align}\label{eq:I_gg^13 2nd}
I_{\gamma\gamma}^{1,3}=\frac{1}{2}I^{\rm u}_{\gamma\gamma}+s_\mu^1 s_\nu^3\Bigg\{&-\frac{1}{2}G_M^2\frac{m_l^2}{\tau m_N^2} g^{\mu\nu}+2\frac{G_E^2+\tau G_M^2}{1+\tau}\Bigg[\nonumber\\
&\frac{g^{\mu\nu}}{4}\Bigg(R^2-\frac{1+\tau}{\tau}\Bigg)+\frac{q^\mu q^\nu}{2\tau q^2}-\frac{p_2^{\{\mu} p_4^{\nu\}}+R q_{\phantom{1}}^{[\mu} p_2^{\nu]}}{q^2}\Bigg]\Bigg\}
\end{align}
is useful to show that $G_M^2$ and $G_E^2$ are separated, as unpolarized case, in spin asymmetry part. By choosing kinematics conditions, we can separately measure $G_E^2$ or $G_M^2$. Further discussions are in section \ref{sec:form factor 13}.

\subsubsection{Incoming and Outgoing Nucleons Polarized}
For polarized incoming and outgoing nucleons,
\begin{align}
I^{2,4}=\left(\frac{1}{2}\sum_{z_1}\right)\sum_{z_3}I^{1,2,3,4}=\frac{1}{2}(1+s^2\cdotp s^4)I^{\rm u}+s_\mu^2 s_\nu^4(J^{2,4})^{\mu\nu}.
\end{align}
The results are
\begin{align}
(J_{\gamma\gamma}^{2,4})^{\mu\nu}=&-G_M^2\Bigg(g^{\mu\nu}-\frac{m_l^2}{\tau m_N^2}\frac{q^\mu q^\nu}{q^2}+2\frac{p_1^\mu p_1^\nu+p_3^\mu p_3^\nu}{q^2}\Bigg)\nonumber\\
&-F_2^2\tau\left(R^2-\frac{1+\tau}{\tau}\right)\frac{q^\mu q^\nu}{q^2}+2G_M F_1\frac{q^\mu q^\nu+Rq_{\phantom{1}}^{[\mu} p_1^{\nu]}}{q^2}\\
(J_{\gamma\phi}^{2,4})^{\mu\nu}=&-2\frac{m_l}{m_N}\Bigg(F_2 R\frac{q^\mu q^\nu}{q^2}+\frac{G_M}{\tau}\frac{q_{\phantom{1}}^{[\mu} p_1^{\nu]}}{q^2}\Bigg)\\
(J_{\phi\phi}^{2,4})^{\mu\nu}=&-\left(1+\frac{m_l^2}{\tau m_N^2}\right)\frac{q^\mu q^\nu}{q^2}.
\end{align}
The asymmetry part of $I_{\gamma\gamma}^{2,4}$ contains the $G_M G_E$ cross term, therefore it is harder to use it to measure the form factors separately.

\subsection{Fully Polarized\label{sec:fully polarized}}
From the argument at the beginning of section \ref{sec:partially polarized}, the fully polarized cross section depends on the product of even number of spin polarizations. One can separate the unpolarized and partially polarized contributions, then leave the part which depends on all four polarizations,
\begin{align}
I_{\gamma\gamma}^{1,2,3,4}=&-\frac{1}{4}\left[5+s^1\cdotp s^3+s^2\cdotp s^4+(s^1\cdotp s^3)(s^2\cdotp s^4)\right]I^{\rm u}_{\gamma\gamma}\nonumber\\
&+\frac{1}{2}\left[(1+s^2\cdotp s^4)I_{\gamma\gamma}^{1,3}+(1+s^1\cdotp s^3)I_{\gamma\gamma}^{2,4}\right]+\frac{1}{4}I_{\gamma\gamma}^{1,2}+\frac{1}{2}(I_{\gamma\gamma}^{1,4}+I_{\gamma\gamma}^{3,2})+I_{\gamma\gamma}^{3,4}\nonumber\\
&+2G_M^2s^1_\mu s^2_\nu s^3_\rho s^4_\sigma\Bigg\{-\frac{F_2^2}{G_M^2}\frac{q^\nu q^\sigma}{q^2}\Bigg(\frac{q^\mu q^\rho}{2q^2}-\tau\frac{p_2^{\{\mu} p_4^{\rho\}}+Rq_{\phantom{1}}^{[\mu} p_2^{\rho]}}{q^2}\Bigg)\nonumber\\
&-\frac{F_1}{G_M}\Bigg[\frac{(p_2+p_4)^{\{\mu} g^{\rho\}[\nu}q^{\sigma]}+Rq^{[\mu}g^{\rho][\nu}q^{\sigma]}}{4q^2}+\frac{q^\mu q^\nu q^\rho q^\sigma+q_{\phantom{1}}^{[\mu} p_2^{\rho]}q_{\phantom{1}}^{[\nu} p_1^{\sigma]}}{q^4}\Bigg]\nonumber\\
&+\frac{g^{\mu\nu}g^{\rho\sigma}+g^{\mu\sigma}g^{\nu\rho}}{4}+\frac{q^{[\mu}g^{\sigma][\nu}q^{\rho]}+g^{\nu[\mu}q^{\rho]}p_1^\sigma+g^{\sigma[\mu}q^{\rho]}p_3^\nu}{2q^2}\Bigg\}\label{eq:I_gg fully}\\
I_{\gamma\phi}^{1,2,3,4}=&-I^{\rm u}_{\gamma\phi}+\frac{I_{\gamma\phi}^{1,3}I_{\gamma\phi}^{2,4}}{I^{\rm u}_{\gamma\phi}}+\frac{1}{4}I_{\gamma\phi}^{1,2}+\frac{1}{2}(I_{\gamma\phi}^{1,4}+I_{\gamma\phi}^{3,2})+I_{\gamma\phi}^{3,4}\nonumber\\
&-s^1_\mu s^2_\nu s^3_\rho s^4_\sigma G_M\frac{m_l}{\tau m_N}\Bigg(\frac{q^{[\rho} g^{\mu][\nu}q^{\sigma]}}{2q^2}+\frac{2}{R}\frac{q_{\phantom{1}}^{[\mu} p_2^{\rho]}q_{\phantom{1}}^{[\nu}p_1^{\sigma]}}{q^4}\Bigg)\\
I_{\phi\phi}^{1,2,3,4}=&\frac{I_{\phi\phi}^{1,3}I_{\phi\phi}^{2,4}}{I^{\rm u}_{\phi\phi}}.
\end{align}

\subsection{Massless Lepton Limit}\label{sec:massless lepton}
As we discussed in section \ref{sec:polarization of massless particle}, naively taking the limit $m_l\rightarrow 0$ does not yield correct results. Only some results are still safe when we take the limit $m_l\rightarrow 0$, such as $I^{\rm u}$, $I^{i,j}_{\gamma\gamma}$, $I^{2,4}$, etc. We discuss the general massless lepton limit in this section.

For a lepton of negligible mass, from either direct calculation or conservation of angular momentum, the interference term $I_{\gamma\phi}$ vanishes for all cases. Therefore it is sufficient to display $I_{\gamma\gamma}$ and $I_{\phi\phi}$.

\begin{enumerate}
\item
The unpolarized $I^{\rm u}$ is safe in the limit $m_l\rightarrow0$,
\begin{align}
&I^{\rm u}_{\gamma\gamma}=\frac{G_E^2+\tau G_M^2}{1+\tau}\left(R^2-\frac{1+\tau}{\tau}\right)+2G_M^2\\
&I^{\rm u}_{\phi\phi}=\frac{1+\tau}{\tau}.
\end{align}

\item For one lepton and one nucleon polarized,
\begin{align}
I^{i,j}=a^{i,j}\sum_{z_{k}}\sum_{z_{l}}I^{i,j,k,l}=a^{i,j}\Big[I^{\rm u}-z^i s_\mu^j (J^j)^\mu\Big]
\end{align}
the result can also be obtained by setting the lepton polarization to be longitudinal $s^i_\mu\rightarrow p^i_\mu/m_l$ then taking $m_l\rightarrow0$ limit,
\begin{align}
(J_{\gamma\gamma}^j)^\mu=&G_M\Bigg[b^j F_2 R\frac{q^\mu}{m_N}-\frac{G_E}{\tau}\frac{(p_1+p_3)^\mu}{m_N}\Bigg]\\
(J_{\phi\phi}^j)^\mu=&0.
\end{align}

\item For incoming and outgoing leptons polarized, one can not directly take the limit $m_l\rightarrow0$, but the results are
\begin{align}
I_{\gamma\gamma}^{1,3}=&\frac{1}{2}(1+z^1z^3)I_{\gamma\gamma}^{\rm u}\\
I_{\phi\phi}^{1,3}=&\frac{1}{2}(1-z^1z^3)I_{\phi\phi}^{\rm u}.
\end{align}

\item For incoming and outgoing nucleons polarized, it is safe to take the limit $m_l\rightarrow0$,
\begin{align}
I^{2,4}=\left(\frac{1}{2}\sum_{z_1}\right)\sum_{z_3}I^{1,2,3,4}=\frac{1}{2}(1+s^2\cdotp s^4)I^{\rm u}+s_\mu^2 s_\nu^4(J^{2,4})^{\mu\nu}.
\end{align}
The results are
\begin{align}
(J_{\gamma\gamma}^{2,4})^{\mu\nu}=&-G_M^2\Bigg(g^{\mu\nu}+2\frac{p_1^\mu p_1^\nu+p_3^\mu p_3^\nu}{q^2}\Bigg)\nonumber\\
&-F_2^2\tau\left(R^2-\frac{1+\tau}{\tau}\right)\frac{q^\mu q^\nu}{q^2}+2G_M F_1\frac{q^\mu q^\nu+Rq_{\phantom{1}}^{[\mu} p_1^{\nu]}}{q^2}\\
(J_{\phi\phi}^{2,4})^{\mu\nu}=&-\frac{q^\mu q^\nu}{q^2}.
\end{align}

\item Finally, for fully polarized case, the result is much simpler than (\ref{eq:I_gg fully})
\begin{align}
I_{\gamma\gamma}^{1,2,3,4}=&-I^{\rm u}_{\gamma\gamma}+\frac{1}{2}(1+z^1 z^3)I_{\gamma\gamma}^{2,4}+\frac{1}{4}I_{\gamma\gamma}^{1,2}+\frac{1}{2}(I_{\gamma\gamma}^{1,4}+I_{\gamma\gamma}^{3,2})+I_{\gamma\gamma}^{3,4}\\
I_{\phi\phi}^{1,2,3,4}=&\frac{I_{\phi\phi}^{1,3}I_{\phi\phi}^{2,4}}{I^{\rm u}_{\phi\phi}}.
\end{align}
\end{enumerate}

\section{Searching for a Scalar Boson\label{sec:searching for a scalar boson}}
If we want to see a new physics signal in elastic lepton-nucleon scattering using one photon and one-scalar-boson exchange, it needs to be greater than the next leading order contribution of the standard model, which is the two-photon exchange contribution (generically suppressed by one more fine structure constant $\alpha$ than one-photon exchange). Therefore, the contribution involving scalar boson, $\left(\frac{d\sigma}{d\Omega}\right)_{\gamma\phi}$ and $\left(\frac{d\sigma}{d\Omega}\right)_{\phi\phi}$, must be greater than at least few percent of $\left(\frac{d\sigma}{d\Omega}\right)_{\gamma\gamma}$ to be detected.

We adopt the constraints in \cite{TuckerSmith:2010ra} that for $m_\phi=1$ MeV
\begin{align}
\frac{g_e}{e}\lesssim 2.3\times10^{-4},\;\frac{g_p}{e}=\frac{g_\mu}{e}\lesssim 1.3\times10^{-3},\;g_n\lesssim 2\times10^{-5}.
\end{align}
Therefore, recall (\ref{eq:lambda}), $|\lambda|\lesssim10^{-6}$. In general, $\left(\frac{d\sigma}{d\Omega}\right)_{\gamma\phi}$ and $\left(\frac{d\sigma}{d\Omega}\right)_{\phi\phi}$ are suppressed by $\lambda$ and $\lambda^2$ with respect to the leading standard model contribution $\left(\frac{d\sigma}{d\Omega}\right)_{\gamma\gamma}$, respectively. Although it seems to be hard to observe scalar boson in elastic scattering, we still find some kinematic regions where the scalar boson may be found. This is because for such kinematic region, $\left(\frac{d\sigma}{d\Omega}\right)_{\gamma\gamma}$ goes or approaches to zero, whereas $\left(\frac{d\sigma}{d\Omega}\right)_{\gamma\phi}$ and $\left(\frac{d\sigma}{d\Omega}\right)_{\phi\phi}$ go or approach to zero slower and eventually dominate. Those regions are usually narrow in parameter space and hard to measure due to polar angle measuring range and flux intensity.

\subsection{Unpolarized, One Lepton and One Nucleon Polarized}
\begin{figure}
\centering
\includegraphics[scale=1.2]{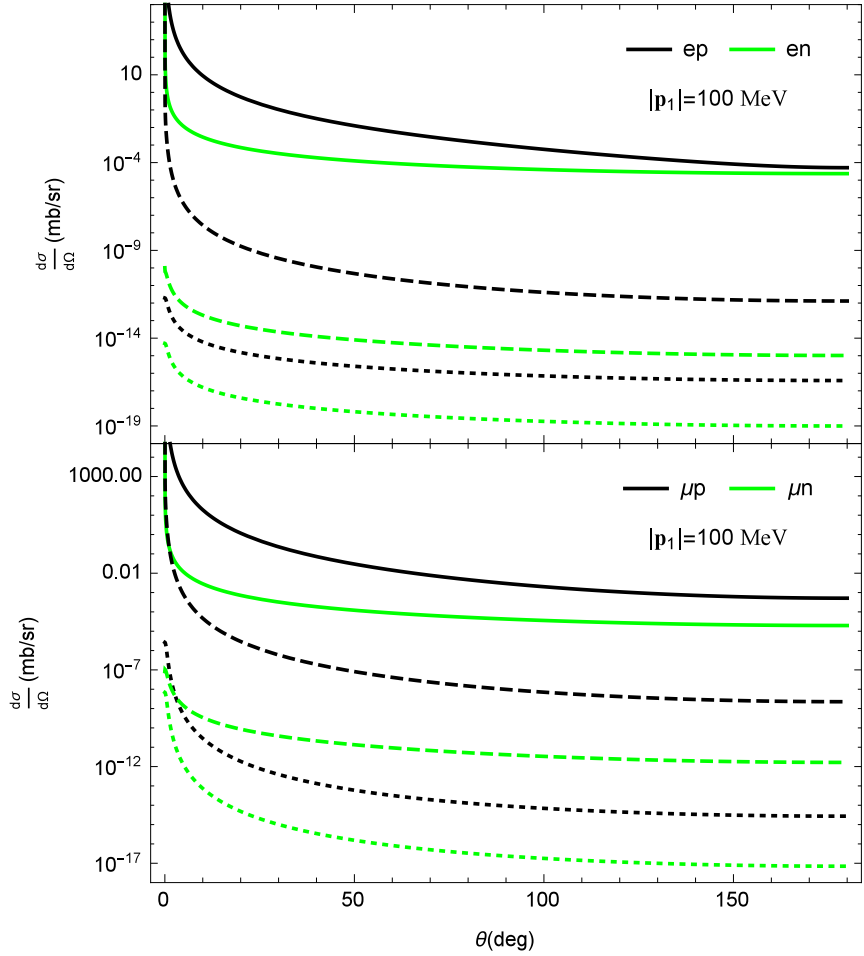}
\caption{\label{fig:unpolarized cross section} (Color Online) Unpolarized differential cross section in the lab frame for $|\mathbf{p}_1|$=100 MeV: the upper (lower) figure is the elastic electron-nucleon (muon-nucleon) scattering cross section; the black (green) line corresponds to proton (neutron); the solid, dashed, and dotted lines correspond to $\left(\frac{d\sigma}{d\Omega}\right)_{\gamma\gamma}$, $\left(\frac{d\sigma}{d\Omega}\right)_{\gamma\phi}$, and $\left(\frac{d\sigma}{d\Omega}\right)_{\phi\phi}$, respectively. It shows that $\left(\frac{d\sigma}{d\Omega}\right)_{\gamma\phi}$ and $\left(\frac{d\sigma}{d\Omega}\right)_{\phi\phi}$ are in general suppressed by $|\lambda|\lesssim10^{-6}$ and $|\lambda^2|\lesssim10^{-12}$, respectively. $\left(\frac{d\sigma}{d\Omega}\right)_{\gamma\phi}$ with muon is greater about two order of magnitudes than $\left(\frac{d\sigma}{d\Omega}\right)_{\gamma\phi}$ with electron, and it can be understood by $\left(\frac{d\sigma}{d\Omega}\right)_{\gamma\phi}$ containing $\frac{m_l}{m_N}$ factor in (\ref{eq:I_gp unpol}). Note that there can be an overall minus sign in front of $\left(\frac{d\sigma}{d\Omega}\right)_{\gamma\phi}$ depending on the sign of $\lambda$. $\left(\frac{d\sigma}{d\Omega}\right)_{\gamma\gamma}$ and $\left(\frac{d\sigma}{d\Omega}\right)_{\phi\phi}$ are always greater or equal to zero.}
\end{figure}

The unpolarized differential cross section for $|\mathbf{p}_1|$=100 MeV is shown in figure \ref{fig:unpolarized cross section}. One can see that generally $\left(\frac{d\sigma}{d\Omega}\right)_{\gamma\phi}$ and $\left(\frac{d\sigma}{d\Omega}\right)_{\phi\phi}$ are suppressed by $|\lambda|\lesssim10^{-6}$ and $|\lambda^2|\lesssim10^{-12}$, respectively. Also, there is about two orders of magnitude enhancement in $\left(\frac{d\sigma}{d\Omega}\right)_{\gamma\phi}$ if the lepton is muon instead of electron. This can be understood by $\left(\frac{d\sigma}{d\Omega}\right)_{\gamma\phi}$ containing $\frac{m_l}{m_N}$ factor in (\ref{eq:I_gp unpol}). We do not observe any kinematic region, which depends only on $\theta$ for the unpolarized case, for the scalar boson to be observable.

For one lepton and one nucleon polarized, by examining the parameter space $(\theta,\alpha_i,\beta_i,\alpha_j,\beta_j)$, we do not find any kinematic region where the effect of the scalar boson is significant either.

\subsection{Incoming and Outgoing Leptons Polarized\label{sec:searching for phi_leptons polarized}}
\begin{figure}
\centering
\subfigure[$|\mathbf{p}_1|=100$ MeV]{
\includegraphics[scale=1.2]{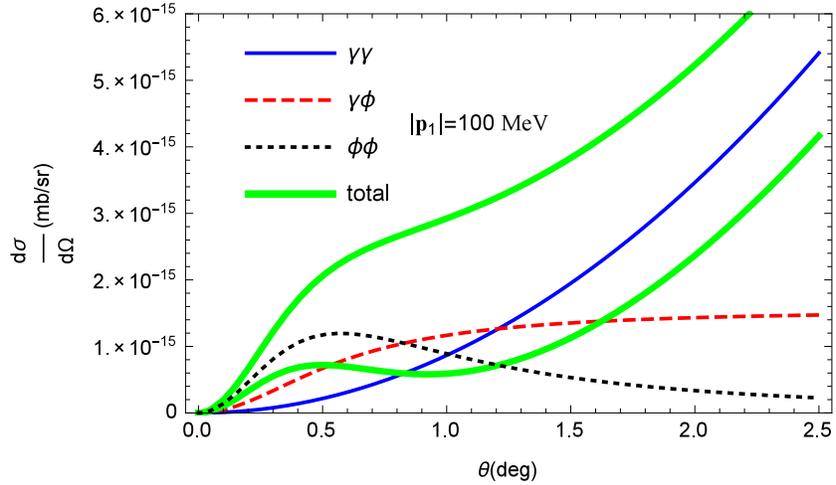}}
\subfigure[$|\mathbf{p}_1|=1$ GeV]{
\includegraphics[scale=1.2]{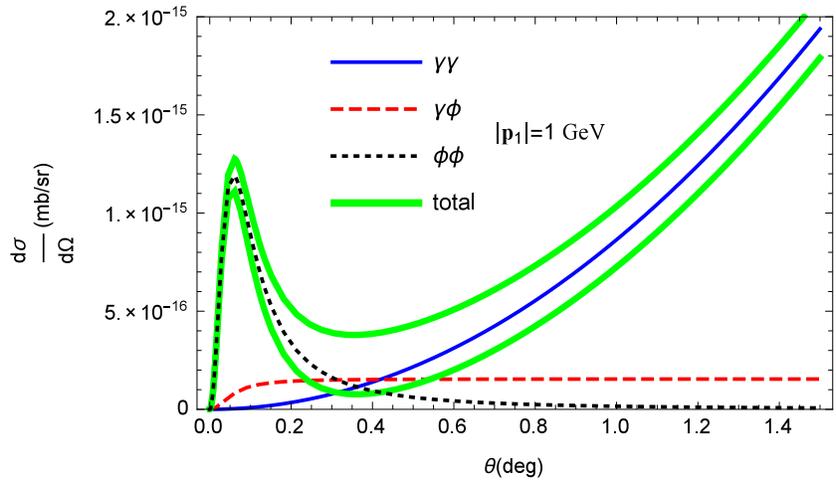}}
\caption{\label{fig:cross_section_13} (Color Online) Elastic electron-neutron forward scattering with electron helicity flip ($\alpha_1=0$, $\alpha_3=180^\circ$, and $\theta\rightarrow0$) in the lab frame: the blue, dashed red, dotted black, and thick green lines correspond to $\left(\frac{d\sigma}{d\Omega}\right)_{\gamma\gamma}$, $\left(\frac{d\sigma}{d\Omega}\right)_{\gamma\phi}$, $\left(\frac{d\sigma}{d\Omega}\right)_{\phi\phi}$, and total differential cross section, respectively. There can be an overall minus sign in front of $\left(\frac{d\sigma}{d\Omega}\right)_{\gamma\phi}$ depending on the sign of $\lambda$; based on the same reason, there are two green lines depending on the choice of particle or anti-particle on electron and neutron.}
\end{figure}

The differential cross section of polarized incoming and outgoing leptons in the lab frame may be good for finding scalar boson in electron-neutron scattering. For helicity flip forward scattering ($\alpha_1=0$, $\alpha_3=180^\circ$, and $\theta\rightarrow0$), although the differential cross section goes to zero, $\left(\frac{d\sigma}{d\Omega}\right)_{\gamma\phi}$ and $\left(\frac{d\sigma}{d\Omega}\right)_{\phi\phi}$ go to zero much slower than $\left(\frac{d\sigma}{d\Omega}\right)_{\gamma\gamma}$ and dominate at small angle. See figure \ref{fig:cross_section_13}, we can look for a bump to observe the scalar boson. One can find that $|\mathbf{p}_1|$ around 100 MeV may be a good choice. If $|\mathbf{p}_1|$ is smaller, $\left(\frac{d\sigma}{d\Omega}\right)_{\gamma\gamma}$ dominates; if $|\mathbf{p}_1|$ is bigger, the region where $\left(\frac{d\sigma}{d\Omega}\right)_{\gamma\phi}$ and $\left(\frac{d\sigma}{d\Omega}\right)_{\phi\phi}$ dominate is closer to $\theta=0$ and the region where scalar contribution dominates becomes narrower.

One facility to measure electron nucleon elastic scattering is the SoLID (Solenoidal Large Intensity Device) in Hall A at JLab \cite{SoLID:2013fta}. The polar angle resolution is around 1 mr, so the precision is sufficient to find the bump. To estimate the production rate, the unpolarized luminosity is $1.3\times10^{39}$ $\rm cm^{-2}s^{-1}$ for $\rm LD_2$ target; the differential cross section is around $10^{-15}$ mb/sr for the scattering angle from $0^\circ$ to $2^\circ$ (figure \ref{fig:cross_section_13} for $|\mathbf{p}_1|=100$ MeV). Combining everything above, it will be about 2 days for an event to occur.

Although the number seems promising, the experiment faces several obstacles: the polar angle is outside the minimum equipment coverage range which is about $8^\circ$; polarizing the incoming electron reduces the luminosity; measuring the polarization of the the outgoing electron is very difficult. Moreover, because the target is liquid deuterium, to separate electron-neutron scattering signals from much bigger electron-proton scattering signals is difficult. The differential cross section for electron-proton scattering blows up in the forward direction, and it is about $10^{-4}$ ($10^{-2}$) mb/sr around $1^\circ$ ($0.1^\circ$).

\subsection{Incoming and Outgoing Nucleons Polarized\label{sec:searching for phi_nucleons polarized}}
\begin{figure}
\centering
\includegraphics[scale=1.2]{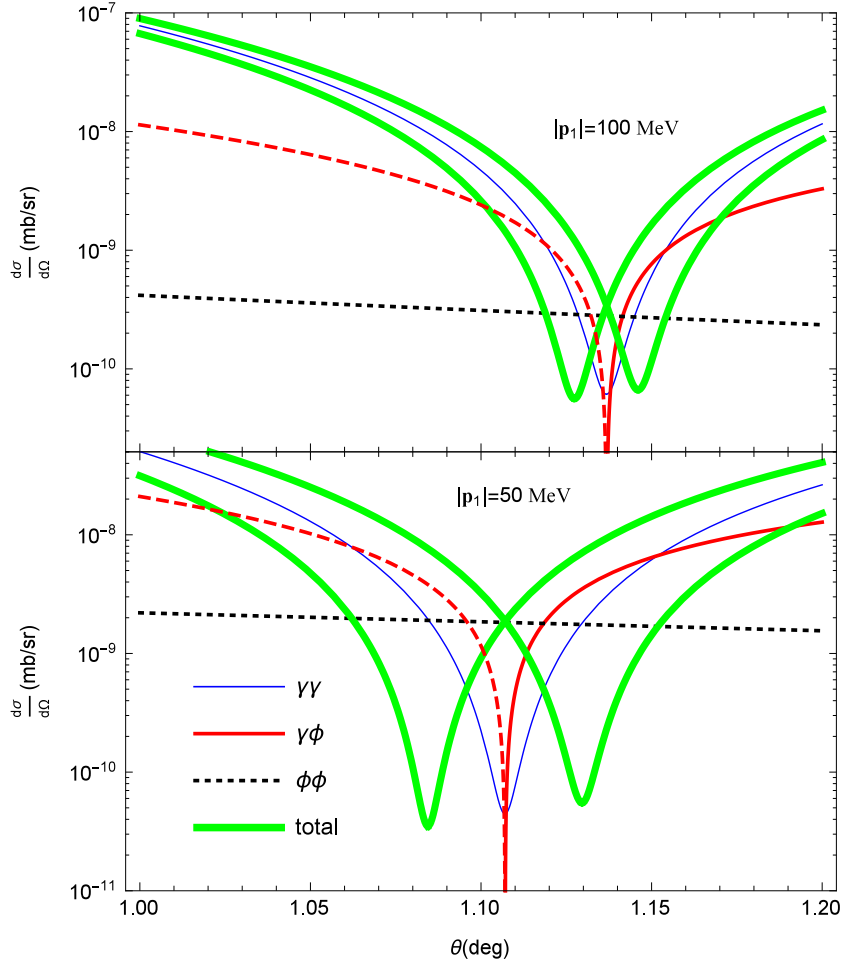}
\caption{\label{fig:cross_section_24} (Color Online) Elastic muon-neutron scattering for incoming and outgoing neutrons polarized $(\alpha_2,\beta_2,\alpha_4,\beta_4)=(90.5^\circ,0^\circ,180^\circ,0^\circ)$ in the lab frame: the thin blue, (solid and dashed) red, dotted black, and thick green lines correspond to $\left(\frac{d\sigma}{d\Omega}\right)_{\gamma\gamma}$, $\left(\frac{d\sigma}{d\Omega}\right)_{\gamma\phi}$, $\left(\frac{d\sigma}{d\Omega}\right)_{\phi\phi}$, and total differential cross section, respectively. The solid and dashed red lines means they are differed by a minus sign, and there can be an overall minus sign in front of $\left(\frac{d\sigma}{d\Omega}\right)_{\gamma\phi}$ depending on the sign of $\lambda$; based on the same reason, there are two green lines depending on the choice of particle or anti-particle on muon and neutron.}
\end{figure}

The differential cross section of polarized incoming and outgoing nucleons in the lab frame may be good for finding scalar boson in muon-neutron scattering in neutron rest frame. For the spin polarization in the scattering plane ($\beta_2=0$ and $\beta_4=0$), one can choose $\alpha_2=90.5^\circ$ and $\alpha_4=180^\circ$, then there is a region near $0^\circ$ (depending on $|\mathbf{p}_1|$) where the contribution of $\left(\frac{d\sigma}{d\Omega}\right)_{\gamma\phi}$, and $\left(\frac{d\sigma}{d\Omega}\right)_{\phi\phi}$ dominates. See figure \ref{fig:cross_section_24}, we can look for the shift of the local minimum predicted by standard model. One can find that $|\mathbf{p}_1|$ around 50 MeV may be a good choice. If $|\mathbf{p}_1|$ is smaller, the region which $\left(\frac{d\sigma}{d\Omega}\right)_{\gamma\phi}$ and $\left(\frac{d\sigma}{d\Omega}\right)_{\phi\phi}$ dominate is closer to $\theta=0$; if $|\mathbf{p}_1|$ is bigger, $\left(\frac{d\sigma}{d\Omega}\right)_{\gamma\gamma}$ dominates.

For MUSE at PSI \cite{MUSE}, the polar angle resolution is around 1 mr, which is a sufficient precision to observe the shift of the local minimum if we combine muon-neutron and anti-muon-neutron scattering. To estimate the production rate, the muon unpolarized flux about $2\times10^5$ Hz at momentum 115 MeV; the target $\rm LD_2$ has thickness 4 cm and density 162.4 $\rm kg/m^3$; the differential cross section is around $10^{-9}$ mb/sr for the scattering angle from $1.0^\circ$ to $1.2^\circ$ (figure \ref{fig:cross_section_24} for $|\mathbf{p}_1|=50$ MeV). Combining all numbers above, it will take about 1000 years for an event to occur.

The main obstacles for this experiment is that the intensity of muon flux is much smaller than electron flux. It also suffers from other difficulties: the polar angle is outside the equipment coverage $20^\circ$ to $100^\circ$; because the target is liquid deuterium, to separate muon-neutron scattering signals from much bigger muon-proton scattering signals is difficult (the differential cross section is about $10^6$ mb/sr around $1.1^\circ$).

\section{Measuring $G_E$ and $G_M$\label{sec:measuring form factors}}
In this section, we generalize and discuss the currently used methods \cite{Rock:2001zi,Arrington:2011kb,Punjabi:2014tna}, and examine alternative ways to measure form factors when incoming and outgoing leptons are polarized.

\subsection{Polarize One Lepton and One Nucleon\label{sec:form factor ij}}
From (\ref{eq:I_gg^ij 2nd}), one can choose kinematics conditions such that the spin asymmetry part,
\begin{align}
P^{i,j}=I_{\gamma\gamma}^{i,j}-a^{i,j}I^{\rm u}_{\gamma\gamma}=a^{i,j}s_\mu^i s_\nu^j(J^{i,j})^{\mu\nu},
\end{align}
can selectively receive $G_M^2$ or $G_MG_E$ contributions. In the lab frame, if the spin polarization vector $\mathbf{s}^j$ of nucleon is perpendicular or parallel to momentum transfer vector $\mathbf{q}$, their corresponding contribution to asymmetry part of $I_{\gamma\gamma}^{i,j}$ are $P^{i,j}_T$ and $P^{i,j}_L$,
\begin{align}
P^{i,j}_T=&-2a^{i,j}G_MG_E\frac{m_l}{\tau m_N}s^i\cdotp s^j_T\\
P^{i,j}_L=&2a^{i,j}b^jG_M^2\frac{m_l}{m_N}\frac{s_i\cdotp(p_2+p_4)s^j_L\cdotp q}{|\mathbf{q}|^2},
\end{align}
which leads to the most general from of so-called ratio technique or polarization transfer method to measure the form factors,
\begin{align}\label{eq:ratio tech}
\frac{G_E}{G_M}=-b^j\Bigg(\frac{P_T}{P_L}\Bigg)^{i,j}\frac{\tau}{|\mathbf{q}|^2}\frac{s_i\cdotp(p_2+p_4)s^j_L\cdotp q}{s^i\cdotp s^j_T}.
\end{align}

There are some special cases worth discussing.
\begin{enumerate}
\item
If the lepton is longitudinally polarized,
\begin{align}
\frac{s_i\cdotp(p_2+p_4)s^j_L\cdotp q}{s^i\cdotp s^j_T}=\frac{|\mathbf{q}|}{\hat{\mathbf{p}}_i\cdotp\hat{\mathbf{s}}^j_T}\Big[-2m_N(1+\tau)\frac{|\mathbf{p}_i|}{E_i}+(\mathbf{p}_1-\mathbf{p}_3)\cdotp\hat{\mathbf{p}}_i\Big].
\end{align}

\item
If $\hat{\mathbf{s}}^j_T$ is in scattering plane,
\begin{align}
\hat{\mathbf{p}}_i\cdotp\hat{\mathbf{s}}^j_T=\frac{|\mathbf{p}_1||\mathbf{p}_3|}{|\mathbf{p}_i||\mathbf{q}|}\sin\theta.
\end{align}

\item
If lepton is massless,
\begin{align}
2m_N(1+\tau)\frac{|\mathbf{p}_i|}{E_i}-(\mathbf{p}_1-\mathbf{p}_3)\cdotp\hat{\mathbf{p}}_i\underset{m_l\rightarrow0}{=}\frac{m_N(E_1+E_3)}{E_i}
\end{align}

\item
Combining all the choices and limit above, (\ref{eq:ratio tech}) reduces to a familiar form\footnote{In other literature, $P_T$ and $P_L$ usually stand for spin polarization transfer which is the spin asymmetry part divided by spin symmetry part.}
\begin{align}
\frac{G_E}{G_M}=b^j\Bigg(\frac{P_T}{P_L}\Bigg)^{i,j}\frac{E_1+E_3}{2m_N}\tan\frac{\theta}{2},
\end{align}
which is widely used today to measure form factors in elastic electron-nucleon scattering \cite{Rock:2001zi,Arrington:2011kb,Punjabi:2014tna}.
\end{enumerate}
The commonly used cases are $(i,j)=(1,2)$ or $(1,4)$ because the polarization of outgoing lepton is harder to measure.

\subsection{Incoming and Outgoing Leptons Polarized\label{sec:form factor 13}}
\begin{figure}
\centering
\includegraphics[scale=1.2]{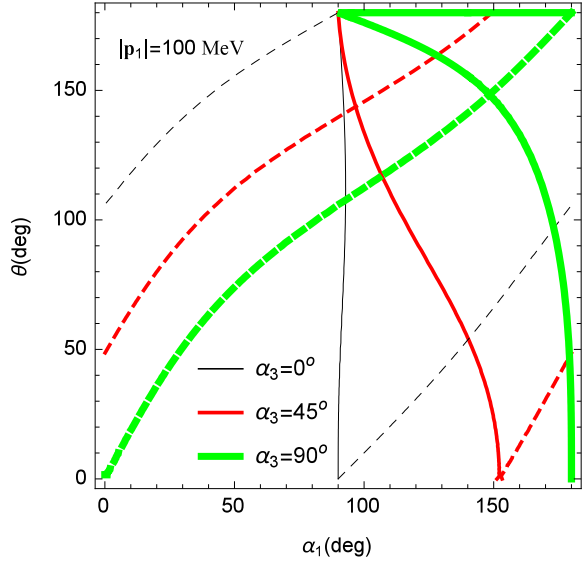}
\caption{\label{fig:P13 zeros} (Color Online) Elastic muon-nucleon scattering for $|\mathbf{p}_1|$=100 MeV with polarized incoming and outgoing muons in the scattering plane, $\beta_1=0$ and $\beta_3=0$, in the lab frame: One can find the zeroes of $G_M^2$ and $G_E^2$ terms in $P^{1,3}$ denoted as the solid and dashed lines in the figure,respectively; the thin black, red, and thick green lines correspond to $\alpha_3=0^\circ$, $45^\circ$, and $90^\circ$, respectively. The zeros for muon-proton and muon-neutron scattering are actually different, but too small to distinguish in the figure.}
\end{figure}

From (\ref{eq:I_gg^13 2nd}), one can choose kinematics conditions such that the spin asymmetry part
\begin{align}
P^{1,3}=I_{\gamma\gamma}^{1,3}-\frac{1}{2}I^{\rm u}_{\gamma\gamma}=\frac{1}{2}s^1\cdotp s^3I+s_\mu^1 s_\nu^3(J^{1,3})^{\mu\nu}
\end{align}
can selectively receive $G_M^2$ or $G_E^2$ contributions. We do not find a compact algebraic form for such condition, but one can numerically evaluate it. In figure \ref{fig:P13 zeros}, assuming $|\mathbf{p}_1|$=100 MeV and the spin polarizations are in scattering plane, $\beta_1=0$ and $\beta_3=0$, we find the zeros of $G_M^2$ and $G_E^2$ terms in $P^{1,3}$ of elastic muon-nucleon scattering.

\begin{figure}
\centering
\includegraphics[scale=1.2]{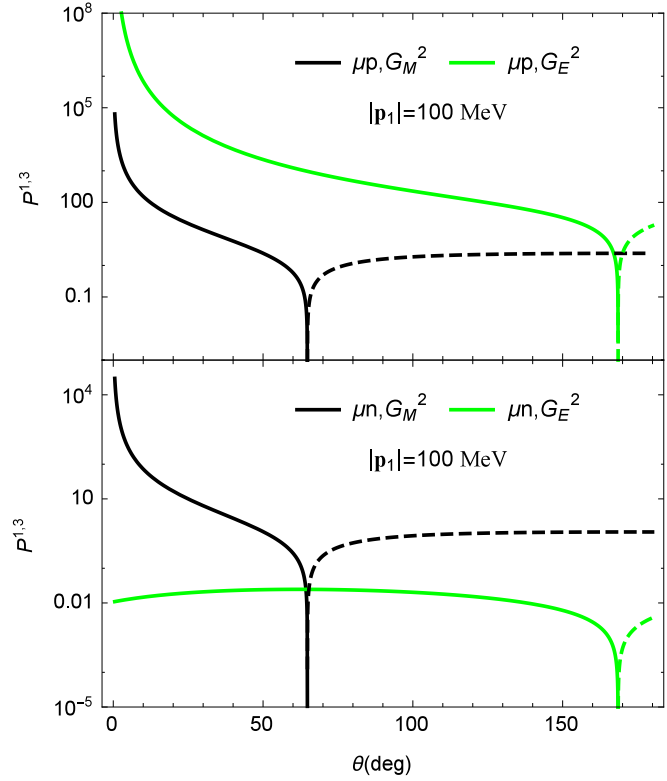}
\caption{\label{fig:P13} (Color Online) Elastic muon-nucleon scattering for $|\mathbf{p}_1|$=100 MeV with polarized incoming and outgoing muons, $(\alpha_1,\beta_1,\alpha_3,\beta_3)=(135^\circ,0^\circ,45^\circ,0^\circ)$: The solid and dashed lines represent the positive and negative values, respectively; the black and green lines correspond to $G_M^2$ and $G_E^2$ terms of $P^{1,3}$ of elastic muon-nucleon scattering cross section. The zeros of $G_M^2$ and $G_E^2$ terms of $P^{1,3}$ for elastic muon-proton and muon-neutron scattering are very close but different.} 
\end{figure}

As an example (see figure \ref{fig:P13}), consider $|\mathbf{p}_1|$=100 MeV and $(\alpha_1,\beta_1,\alpha_3,\beta_3)=(135^\circ,0^\circ,45^\circ,0^\circ)$, for elastic muon-proton scattering, $P^{1,3}$ receives contribution only from $G_E^2$ and $G_M^2$ term at $64.794^\circ$ and $168.508^\circ$, respectively; for elastic muon-neutron scattering, $P^{1,3}$ receives contribution only from $G_E^2$ and $G_M^2$ term at $64.798^\circ$ and $168.506^\circ$, respectively.

In practice, although the angular resolution is sufficient to perform the experiment, however, this method to measure the form factors could be hard due to the fact that to polarized and measure leptons to some specific non-longitudinal angle is not an easy task. 

\section{Conclusions\label{sec:conclusions}}
In this paper, we calculate the differential cross section of lepton-nucleon elastic scattering using the one-photon and one-scalar-boson exchange mechanism for all possible polarizations in a general reference frame. The expressions are shown in section \ref{sec:cross section}. There are some possible applications: finding a new scalar boson to resolve proton radius and muon $g-2$ puzzles; generalizing the current methods to measure the ratio of the nucleon form factors, $G_E/G_M$; providing an alternative way to measure $G_E^2$ and $G_M^2$ directly. 

The effects of new scalar boson are studied in section \ref{sec:searching for a scalar boson}. We conclude that there are two cases that the scalar boson is potentially observable. For elastic electron-neutron helicity flip forward scattering with incoming and outgoing electrons polarized, the effects of scalar boson are dominant and produce a bump, see figure \ref{fig:cross_section_13}. The optimistic estimates in section \ref{sec:searching for phi_leptons polarized} show that the experiment is possible for current technology but faces several obstacles. For elastic muon-neutron scattering at small scattering angles using polarized incoming and outgoing neutrons, the effects of scalar boson is significant and shifts the local minimum, see figure \ref{fig:cross_section_24}. However, the optimistic estimates in section \ref{sec:searching for phi_nucleons polarized} show that unless the intensity of muon flux can be dramatically increased, the experiment is impossible in the foreseeable future.

In section \ref{sec:form factor ij}, the current method to measure nucleon form factors, as known as ratio technique or polarization transfer method, is generalized and shown in (\ref{eq:ratio tech}). The current method is to polarize incoming electron and either polarize incoming or outgoing nucleons. We generalize it such that one of the incoming and outgoing leptons is polarized, and one of the incoming and outgoing nucelons is polarized. Also, in order to separate the contributions of $G_MG_E$ and $G_M^2$ terms, the current method requires polarization vector of the nucleon $\mathbf{s}^j$ to be either perpendicular or parallel to momentum transfer $\mathbf{q}$, and neglects the lepton mass and as a result the lepton polarization become longitudinal. In our expression, we only require the kinematic condition that the nucleon $\mathbf{s}^j$ is either perpendicular or parallel to momentum transfer $\mathbf{q}$. In conclusion, using our new expression, one can choose which leptons and nucleons to be polarized; the lepton mass is included; the polarization of leptons becomes two extra angular parameters in an experiment.

In section \ref{sec:form factor 13}, in studying elastic muon-nucleon scattering with incoming and outgoing muons polarized, we can separately measure the contributions of $G_E^2$ and $G_M^2$. Although we do not obtain a simple analytic expression, it is easy to evaluate the zeros of $G_E^2$ and $G_M^2$ terms of spin asymmetry part of the cross section numerically, see figure \ref{fig:P13 zeros}. As an example in figure \ref{fig:P13}, one can measure $G_E^2$ and $G_M^2$ directly, and the angle resolution of the current facility is sufficient to perform the experiment. However, measuring the polarization of the outgoing muon could be challenging.

\section*{Acknowledgment}
We thank Bridget Bertoni, Seyda Ipek, David McKeen, and Laurence G. Yaffe for valuable suggestions and discussions. The work of G. A. M. and Y.-S. L. were supported by the U. S. Department of Energy Office of Science, Office of Nuclear Physics under Award Number DE-FG02-97ER-41014.

\end{document}